\documentclass[aps,prl,twocolumn,showpacs,preprintnumbers,amsmath,amssymb,superscriptaddress]{revtex4}

\usepackage{graphicx}
\usepackage{dcolumn}
\usepackage{bm}
\usepackage{color}

\begin{document}
\preprint{PREPRINT (\today)}

\title{ Superfluid Density and Energy Gap-Function of Superconducting
PrPt$_4$Ge$_{12}$}
\author{A.~Maisuradze}
 \email{alexander.maisuradze@psi.ch}
\affiliation{Laboratory for Muon Spin Spectroscopy, Paul Scherrer Institut, CH-5232 Villigen PSI, Switzerland}
\author{M.~Nicklas}
\affiliation{Max-Planck-Institut f\"ur Chemische Physik fester
Stoffe, N\"othnitzer Str. 40, 01187 Dresden, Germany}
\author{R.~Gumeniuk}
\affiliation{Max-Planck-Institut f\"ur Chemische Physik fester
Stoffe, N\"othnitzer Str. 40, 01187 Dresden, Germany}
\author{C.~Baines}
\affiliation{Laboratory for Muon Spin Spectroscopy, Paul Scherrer Institut, CH-5232 Villigen PSI, Switzerland}
\author{W.~Schnelle}
\affiliation{Max-Planck-Institut f\"ur Chemische Physik fester Stoffe, N\"othnitzer Str.
40, 01187 Dresden, Germany}
\author{H.~Rosner}
\affiliation{Max-Planck-Institut f\"ur Chemische Physik fester Stoffe, N\"othnitzer Str.
40, 01187 Dresden, Germany}
\author{A.~Leithe-Jasper}
\affiliation{Max-Planck-Institut f\"ur Chemische Physik fester Stoffe, N\"othnitzer Str.
40, 01187 Dresden, Germany}
\author{Yu.~Grin}
\affiliation{Max-Planck-Institut f\"ur Chemische Physik fester Stoffe, N\"othnitzer Str.
40, 01187 Dresden, Germany}
\author{R.~Khasanov}
 \email{rustem.khasanov@psi.ch}
\affiliation{Laboratory for Muon Spin Spectroscopy, Paul Scherrer
Institut, CH-5232 Villigen PSI, Switzerland}
%

\begin{abstract}

The filled skutterudite superconductor PrPt$_4$Ge$_{12}$ was studied
in muon-spin rotation ($\mu$SR), specific heat, and electrical resistivity
experiments.
The continuous increase of the superfluid density with decreasing
temperature and the dependence of the magnetic penetration depth
$\lambda$ on the magnetic field obtained by means of $\mu$SR, as well as the observation of a $T^3$
dependence of the electronic specific heat establish the presence of
point-like nodes in the superconducting energy gap. The energy
gap was found to be well described by $\Delta =\Delta_0| \hat{k}_x
\pm i \hat{k}_y|$ or $\Delta = \Delta_0(1 - \hat{k}_y^4)$
functional forms, similar to that obtained for another skutterudite
superconductor, PrOs$_4$Sb$_{12}$. The gap to $T_c$ ratios were
estimated to be $\Delta_0/k_BT_c = 2.68(5)$ and 2.29(5), respectively.

\end{abstract}

\maketitle

The filled skutterudite compounds $RM_4Pn_{12}$ ($R=$ rare-earth or
alkaline-earth, $M = $ Fe, Ru, Os, and $Pn = $ P, As, Sb) have
attracted much attention in the recent years. Depending on the
composition they may undergo metal-insulator transitions
\cite{Matsunami05}, show conventional or unconventional
superconductivity \cite{Bauer02,MacLaughlin02, Aoki03}, become
magnetic \cite{Grin05, Sayles08, Schnelle08}, semiconducting
\cite{Nakanishi07}, or exhibit the Kondo effect \cite{Hidaka05} {\it etc}.
One of the most interesting discoveries was
superconducting Pr-filled skutterudites, since within the
traditional mechanism of $s-$wave Cooper pairing the Pr magnetism
would destroy superconductivity. The first compound
demonstrating such unusual behavior was PrOs$_4$Sb$_{12}$
\cite{Bauer02}. The Cooper-pairing mechanism and the
corresponding symmetry of the superconducting energy gap in
PrOs$_4$Sb$_{12}$ are still under debate. Studies of the thermal
conductivity suggested the presence of two distinct superconducting
phases \cite{Izawa03}. Penetration depth experiments were
pointing either to a possible nodal structure of the
gap \cite{Elbert03} or to a gap without nodes
\cite{MacLaughlin02}. A number of experimental techniques such as
scanning tunneling microscopy \cite{Suderow04}, thermal conductivity
\cite{Seyfarth06}, and specific heat measurements
\cite{Sakakibara07} are in agreement with a fully developed
isotropic $s-$wave superconducting gap.

Recently, a new Pr-containing skutterudite superconductor
PrPt$_4$Ge$_{12}$ was discovered \cite{Gumeniuk08}. Specific heat
experiments reveal strongly coupled superconductivity with a
transition temperature $T_c\simeq7.9$~K, factor of 4
higher than that in PrOs$_4$Sb$_{12}$ \cite{Bauer02},
thus making PrPt$_4$Ge$_{12}$ more accessible for e.g.
spectroscopic studies.
Here, we report on a study of PrPt$_4$Ge$_{12}$ by means of muon-spin
rotation ($\mu$SR), specific heat, and electrical resistivity. The
linear increase of the superfluid density ($\rho_s$) with decreasing
temperature, its dependence on the  magnetic field, as well as the
observation of a $T^3$ dependence of the electronic specific heat
document presence of the point-like nodes in the
energy gap.
The temperature dependence of $\rho_s$ was analyzed with various gap
models suggested previously for PrOs$_4$Sb$_{12}$ and found to be
well described by two models of axial symmetry with point-like
nodes: $\Delta = \Delta_0|\hat{k}_x \pm i\hat{k}_y|$ or $\Delta =
\Delta_0(1 - \hat{k}_y^4)$.

The sample preparation procedure of PrPt$_4$Ge$_{12}$ is described
in \cite{Gumeniuk08}.
The transverse field (TF) $\mu$SR experiments were performed at the
$\pi$M3 beam line at the Paul Scherrer Institute (Villigen
Switzerland). The sample was field-cooled from above $T_c$  down to
1.5~K and measured as a function of temperature in a series of fields
ranging from 35~mT to 640~mT. Additional experiments down to
$T\simeq0.03$~K were performed at 75~mT. Typical counting statistics
were $\sim7\times10^6$ positron events per each particular data
point.
Electrical resistivity $R(T,H)$ and specific heat $C(T)$
down to 0.4~K were measured in a commercial system (PPMS, Quantum
Design) using an AC bridge (LR-700, Linear Research) and the HC
option of the PPMS, respectively.

\begin{figure}[htb]
\includegraphics[width=0.85\linewidth]{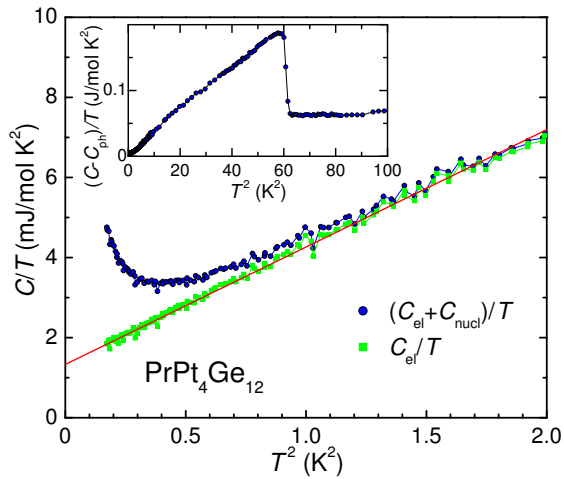}
\caption{(Color online) Inset: specific heat $C-C_{ph}$ of PrPt$_4$Ge$_{12}$ corrected for the
phonon term. Main panel: data below 1.4~K and data corrected
for the nuclear Schottky contribution of Pr (see text).
\label{figCel} }
\end{figure}

Figure~\ref{figCel} shows the electronic specific heat
$C_\mathrm{el}/T$ {\it vs} $T^2$. The phonon contribution $\propto
T^3$ was subtracted using a Debye temperature $\Theta_D$ =
189\,K. The normal state electronic term is
$\gamma_N$ = 63(2) mJ\,mol$^{-1}$\,K$^{-2}$. At the lowest
temperatures an upturn from a nuclear Schottky contribution of Pr
becomes visible ($C_\mathrm{nucl} \propto T^{-2}$). After its
subtraction we find  $C_\mathrm{el}(T) = \gamma' T + \eta T^3$ with
$\gamma' \approx 1.33$ mJ\,mol$^{-1}$\,K$^{-2}$ originating from a
minor metallic impurity phase. For $T/T_c < 0.2$ the clear $T^3$
dependence of the superconducting state electronic term ($\eta = 2.93(2)$~mJ\,mol$^{-1}$\,K$^{-4}$)
suggests that the
gap function of PrPt$_4$Ge$_{12}$ has point-like nodes
\cite{Sigrist91}. The further detailed investigation of the energy
gap and the order parameter symmetry was performed in TF $\mu$SR
experiments.

The TF $\mu$SR data were analyzed as follows. The spatial magnetic
field distribution within the flux line lattice (FLL) was calculated in a
standard way via:
\begin{equation}
B({\bf{r}}) = \langle B \rangle \sum_{\bf{G}} \exp(-i{\bf{G \cdot r}}) B_{\bf{G}}(\lambda, \xi, b).
 \label{eq:Field-distribution}
\end{equation}
Here, $\langle B \rangle$ is the average magnetic field inside the
superconductor, $\xi$ the coherence length, $b = \langle B\rangle/B_{c2}$
the reduced field, $\bf{r}$ the
vector coordinate in a plane perpendicular to the applied field,
${\bf{G}} = 4\pi/\sqrt{3}a(m\sqrt{3}/2, n+m/2)$ the reciprocal
lattice vector of the hexagonal FLL, $a$ the
intervortex distance,  and $m$ and $n$ are integer numbers. The
Fourier components $B_{\bf{G}}$ were obtained by minimizing the
Ginzburg-Landau (GL) free energy using the numerical algorithm
described in Ref.~\onlinecite{Brandt97_NGLmethod}.
The $\mu$SR time-spectra were fit to a theoretical polarization
function $\tilde{P}(t)$ by assuming the internal field distribution
$P_\mathrm{id}(B)$ obtained from Eq.~(\ref{eq:Field-distribution}) and
accounting for the FLL disorder by multiplying $P_\mathrm{id}(B)$ to a
Gaussian function \cite{BrandtJLTPhys88and77}:
\begin{equation}\label{eq:Pt}
\tilde{P}(t) = Ae^{i\phi} \int
e^{-(\sigma_g^2+\sigma_{nm}^2)t^2/2}P_\mathrm{id}(B)e^{i \gamma_{\mu}Bt}dB.
\end{equation}
Here $A$ and $\phi$ are the initial asymmetry and the phase of the
muon spin ensemble, $\sigma_g$ is a parameter related to FLL
disorder \cite{Riesman95} and $\sigma_{nm}$  the nuclear moment
contribution measured at $T>T_c$. For a detailed description of the
fitting procedure we refer to \cite{Maisuradze08}.

\begin{figure}[htb]
\includegraphics[width=0.85\linewidth]{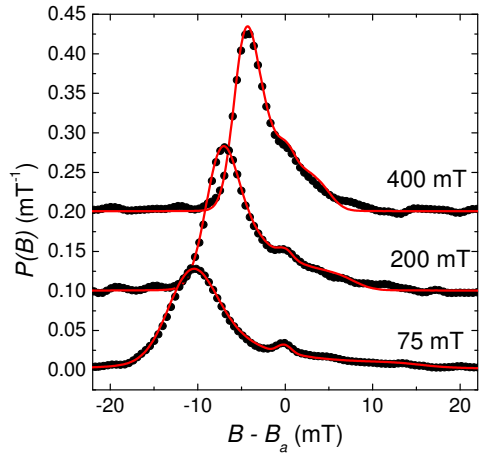}
  \vspace{-0.6cm}
\caption{(Color online) The magnetic field distribution $P(B)$ at
$T=1.55$~K  and applied fields $B_\mathrm{a} = 75$, 200, and 400~mT
obtained by means of fast Fourier transform. The solid lines
are fits to the data (see text for details).}
 \label{fig:Spectra}
\end{figure}

Figure~\ref{fig:Spectra} shows representative internal field
distributions $P(B)$ obtained from the measured $\mu$SR time-spectra
by performing the fast Fourier transform. The solid lines correspond
to the fits by means of the above described procedure. $P(B)$'s have
asymmetric shape as expected for a field distribution within well
arranged FLL.
The small peak in the vicinity of the applied field is
due to muons stopped outside of the sample
($\sim$2-3\%). Above $T_c$ an additional magnetic depolarization is
observed, which increases with increasing field. The measurements at
zero field and above $T_c$ revealed that this magnetic
depolarization is temperature independent.

\begin{figure}[tb]
\begin{tabular}{cc}
{\begin{tabular}{c}
\includegraphics[width=0.55\linewidth,clip]{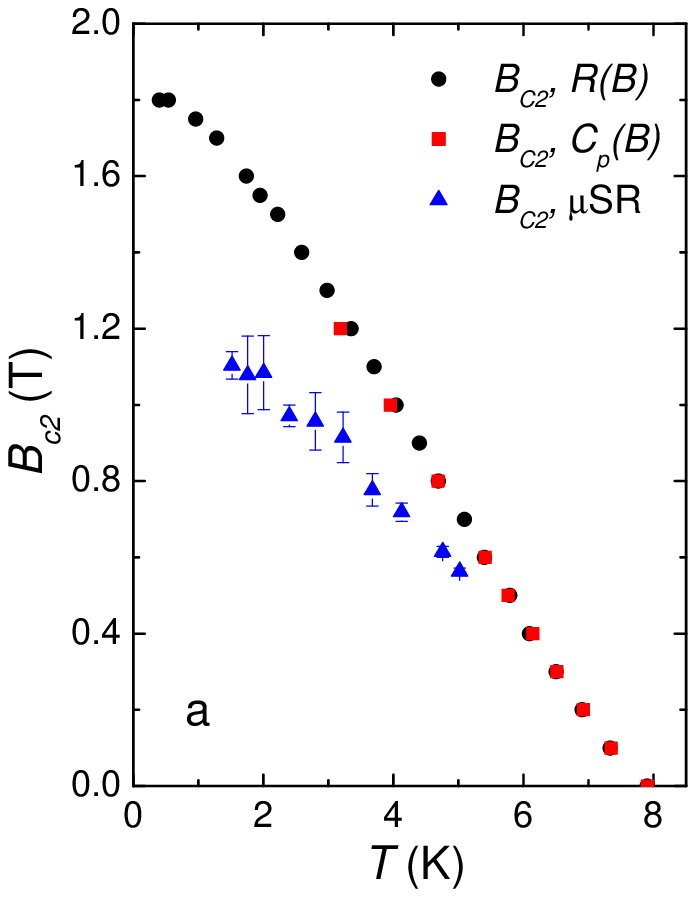} \end{tabular}
}&
{\begin{tabular}{c} \includegraphics[width=0.3\linewidth,clip]{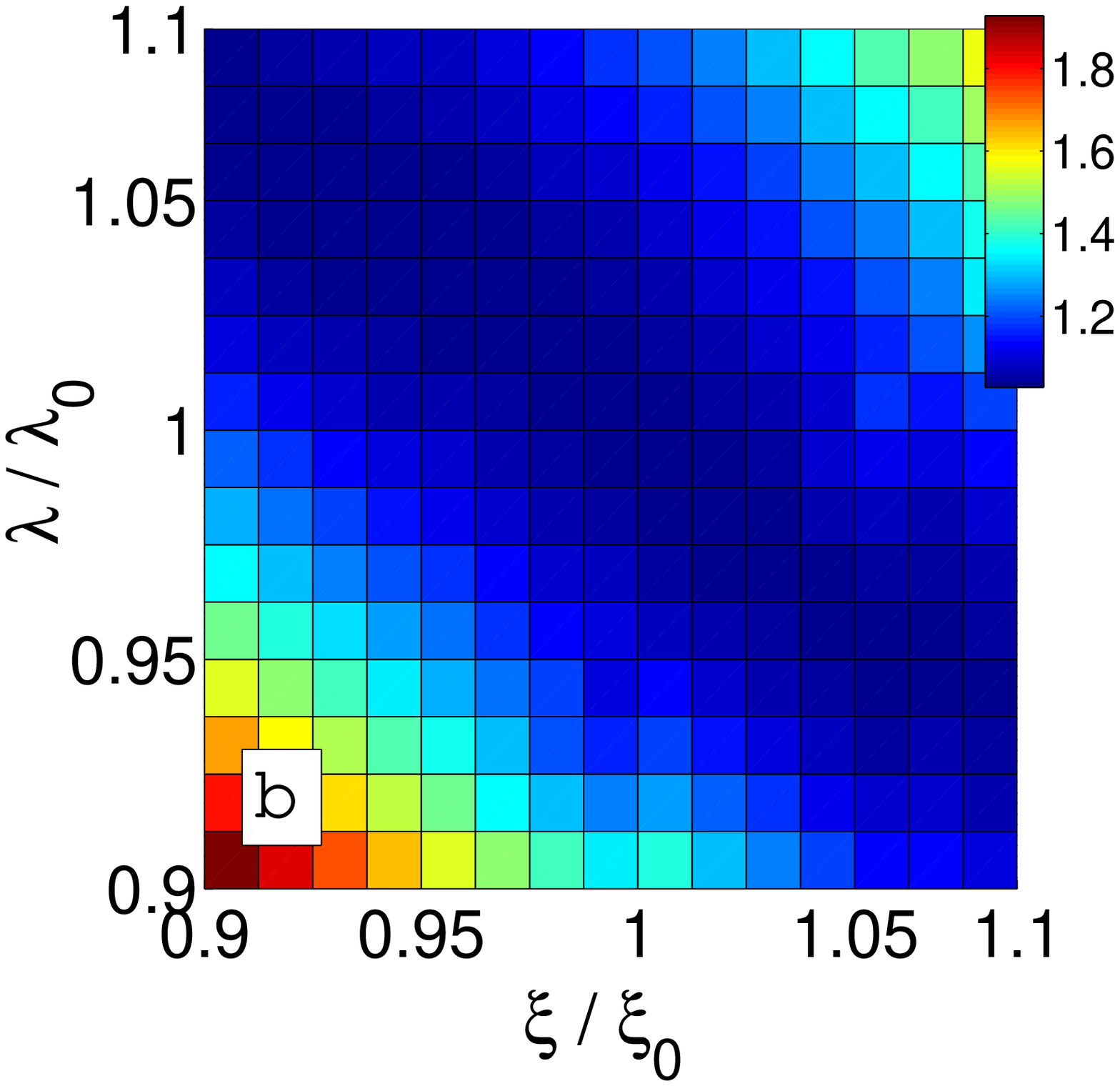}\\
 \includegraphics[width=0.3\linewidth,clip]{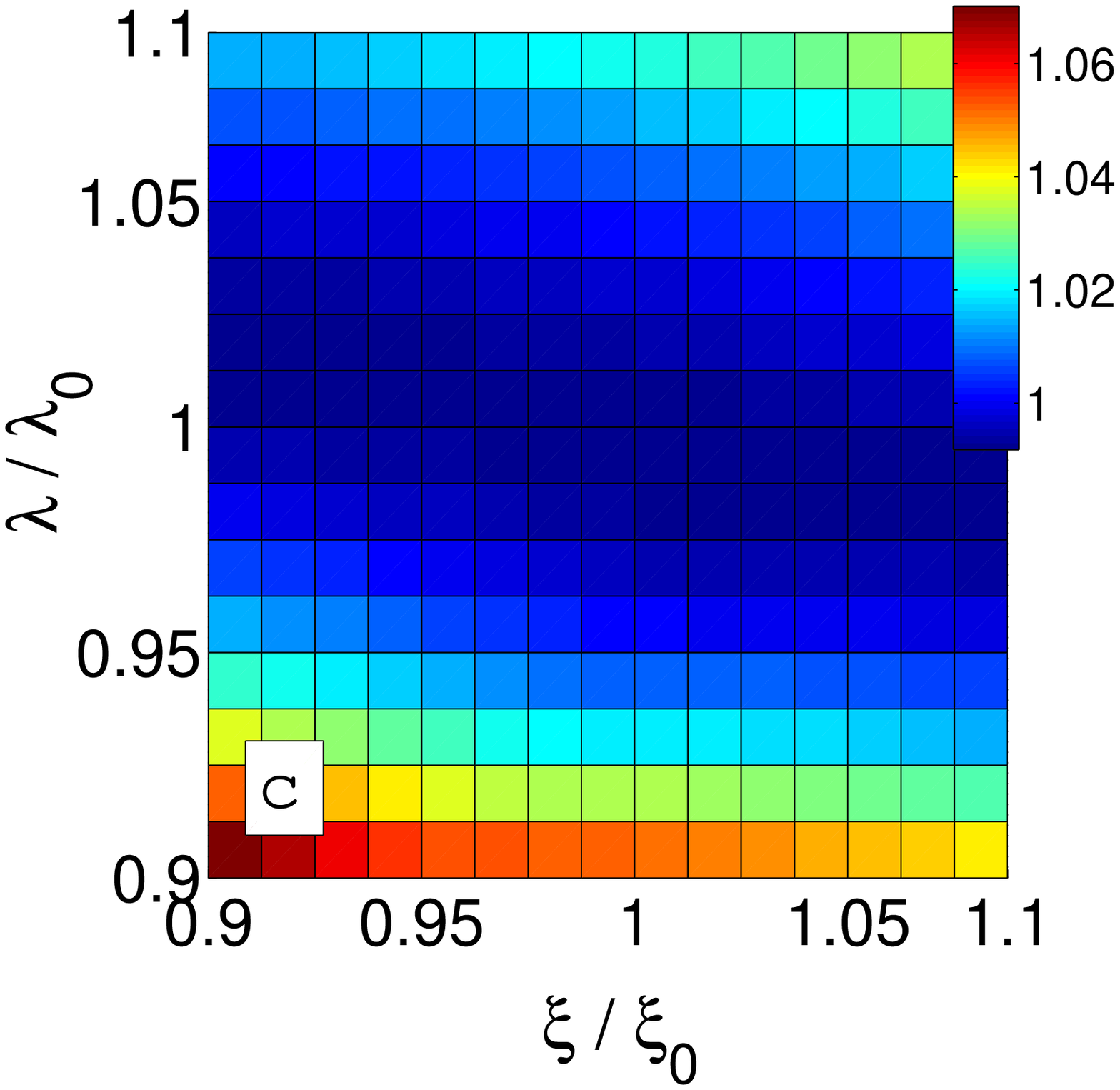}\end{tabular}}
\end{tabular}
  \vspace{-0.6cm}
\caption{(Color online) (a) Temperature dependence of the upper
critical field $B_{c2}$ obtained in specific heat (squares) and
resistivity (dots) experiments  compared with results of the
simultaneous fit of $\mu$SR time spectra at 35 and
400~mT (triangles). Panels (b) and (c) show
correlation plots $\chi^2(\lambda, \xi)$ for $B_\mathrm{a}=400$~mT and
50~mT, respectively.}
 \label{fig:Bc2}
\end{figure}

Magnetic penetration depth is defined in the London limit of low magnetic fields
as a measure of superfluid density $\rho_s \propto 1/\lambda^2$.
The $\lambda$ obtained in the vortex state is often called an effective
magnetic penetration depth $\lambda_{\rm eff}$ \cite{Amin00, LandauKeller07},
since it might be field dependent.
For a classical BCS superconductor with isotropic s-wave gap
$\lambda_{\rm eff} = \lambda$ \cite{LandauKeller07}.
The fit of Eq.~(\ref{eq:Pt}) to the single $\mu$SR
time-spectra does not allow
to obtain independently $\lambda$ and $\xi$, since they
strongly correlate \cite{Maisuradze08}. In order to get rid of this correlation one
needs to perform a {\it simultaneous} fit of several spectra, measured
at the same temperature but different magnetic fields, with
$\lambda$ and $\xi$ as common parameters
\cite{Maisuradze08,Riesman95}. $B_{c2}=\Phi_0/2\pi \xi^2$ as a
function of temperature, obtained after such fit, is presented in
Fig.~\ref{fig:Bc2}a. The fit was performed by combining the data
points measured at the applied fields $B_a=35$ and 400~mT. Figures~\ref{fig:Bc2}b and c
show the corresponding correlation plots [$\chi^2(\lambda,\xi)$].
The normalized $\chi^2$ was obtained for $\lambda$ and $\xi$ varied
within 10\% around their optimal fit values [$\lambda_0 = 114(4)$~nm
and $\xi_0 = 18.1(8)$~nm]. Strong correlations between $\lambda$ and
$\xi$ as well as a field dependent slope of the correlation curves are
obvious.
Figure~\ref{fig:Bc2}a implies, however, that $B_{c2}(T)$ obtained in
the simultaneous fit is approximately 40\% smaller than that
measured directly in specific heat \cite{Gumeniuk08} and resistivity
experiments. This suggests that the assumption of field-independent
$\lambda$ may not be valid for PrPt$_4$Ge$_{12}$ and
$\lambda$ depends on field.
Field dependence of $\lambda$ is expected for a superconductor
with nodes in the gap. A field induces excitations
at the gap nodes due to nonlocal and nonlinear effects, thus
reducing superconducting carrier concentration $n_s \propto 1/\lambda^2$ \cite{Amin00}.
We believe however that GL theory developed for superconductors with
isotropic s-wave gap is approximately valid in the present case.
In order to avoid the
correlation between $\lambda$ and $\xi$ we fixed the values of
$\xi(T)=\sqrt{\Phi_0/2\pi B_{c2}(T)}$ by using the $B_{c2}(T)$ curve
obtained in  resistivity measurements (see Fig.~\ref{fig:Bc2}a) \cite{Gumeniuk08}.

The GL theory allows to extend the London definition of superfluid
density for a finite applied field \cite{Lipavsky02}:

\begin{equation}
\rho_s \propto  \langle |\psi({\bf{r}})|^2\rangle/\lambda^2\simeq(1-b)/\lambda^2.
\label{eq:rho_s}
\end{equation}
Here, $\psi({\bf{r}})$ is the GL order parameter  and
$\langle ...\rangle$ means averaging over the unit cell of the FLL.

\begin{figure}[tb]
\includegraphics[width=0.85\linewidth]{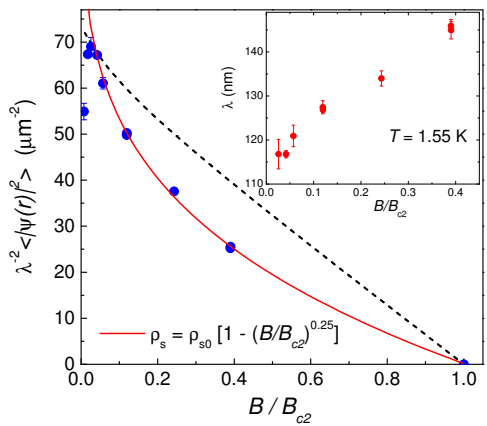}
  \vspace{-0.6cm}
\caption{(Color online) Field dependence of the superfluid density
$\rho_s \propto \langle |\psi|^2 \rangle/\lambda^{2}$. The solid line is the power law
fit $\rho_s(T)/\rho_{s}(T=0)=1-(B/B_{c2})^\alpha$ with $\alpha\simeq0.25$.
The dashed line represents the case when $\lambda$
does not depend on the magnetic field. The inset shows
$\lambda(B/B_{c2})$ at $T = 1.55$~K.}
 \label{fig:SFLdensB}
\end{figure}

Equation~(\ref{eq:rho_s}) implies that in an isotropic s-wave
superconductor the mean superfluid density decreases with
increasing field as $\langle |\psi({\bf{r}})|^2\rangle$
(the dashed line in Fig.~\ref{fig:SFLdensB}).
Obviously,  the experimental $\rho_s$ decreases much stronger with
increasing magnetic field than for a field-independent $\lambda$,
thus suggesting the presence of nodes in the gap.
The power law fit $\rho_s(T)/\rho_{s}(T=0)=1- b^\alpha$ results in
$\alpha = 0.26(2)$ (solid line in Fig.~\ref{fig:SFLdensB}).

\begin{figure}[tb]
\includegraphics[width=0.85\linewidth]{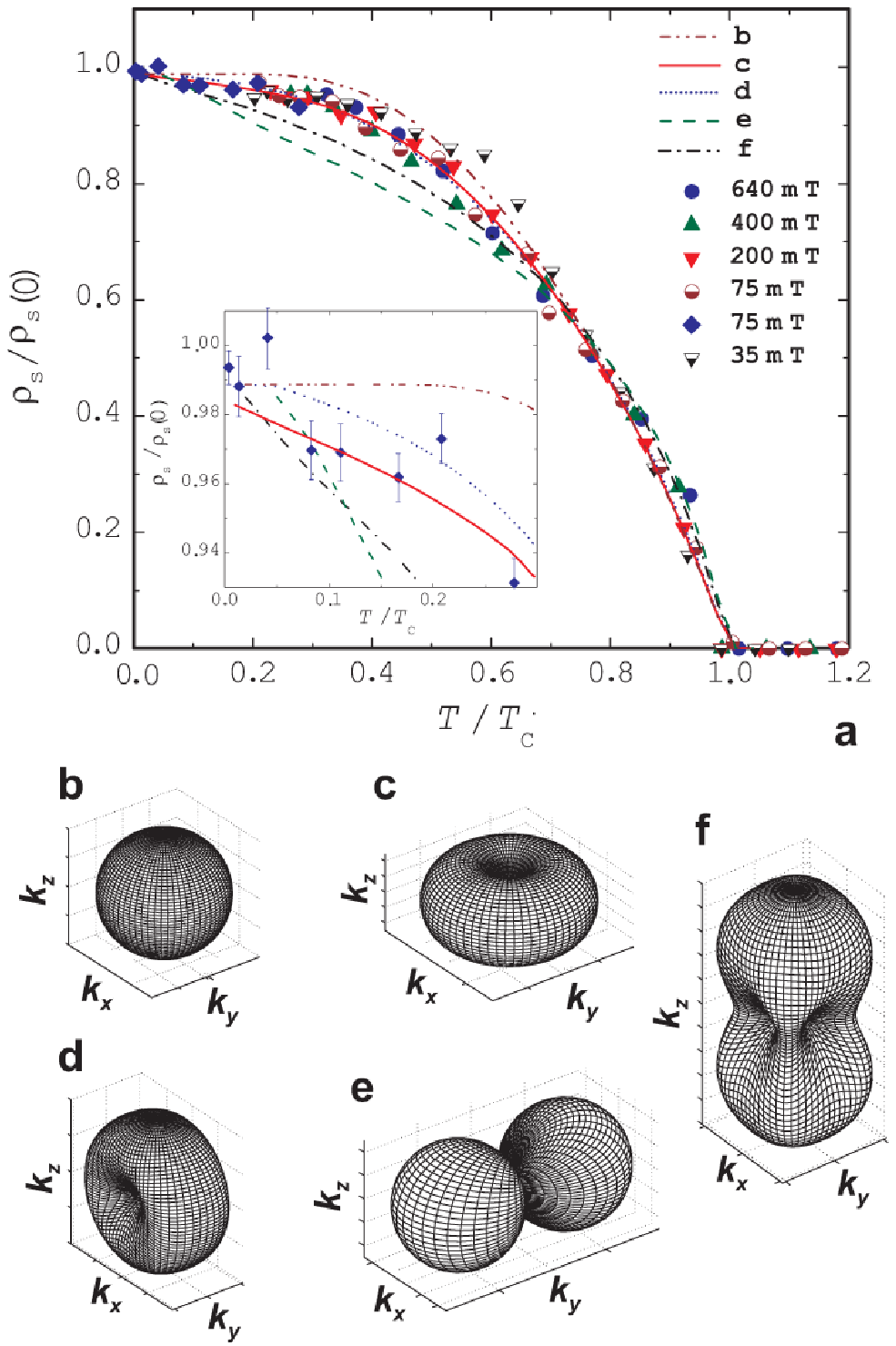}
  \vspace{0.1cm}
\caption{(a) Temperature dependence of $\rho_{s} \propto \langle
|\psi({\bf{r}})|^2\rangle/\lambda^2$ normalized to its value at
$T=30$~mK. The lines represent the fits by using 5 different
gap models.  The inset shows the low
temperature region between $T=0$ and $T=0.3T_c$. The panels from (b)
to (f) represent the gap functions used for the analysis of $\rho_s(T)$
data (see Table \ref{tab:aSFLt} for details). }
 \label{fig:SFLdens_gaps}
\end{figure}

Figure~\ref{fig:SFLdens_gaps}a shows $\rho_s$, normalized to its
value at $T=30$~mK,  as a function of the reduced temperature $t =
T/T_c$. Surprisingly, the $t$ dependence of $\rho_{s}$ is identical for all
applied fields. Only the data points measured at 35~mT deviate
slightly. This may be due to larger vortex disorder and,
consequently, enlarged systematic errors at such a low field.
The inset of Fig.~\ref{fig:SFLdens_gaps}a shows $\rho_s$ measured
at 75~mT for $T\leq0.3T_c$. It is obvious that
$\rho_s$ does not saturate but increases continuously with
decreasing temperature as expected for a superconductor with nodes
in the gap. Note that the presence of nodes is also consistent with
the conclusion drawn from the analysis of $\rho_s(b)$ (see
Fig.~\ref{fig:SFLdensB} and the discussion above).

\begin{table}[htb]
\caption[~]{\label{tab:aSFLt} Summary of the gap analysis of
$\rho_s(T)$ for PrPt$_4$Ge$_{12}$ (see text for details). }
\begin{center}
\begin{tabular}{ccccc}
\hline \hline
 $\Delta(\phi, \theta)$ & Panel in  & $\Delta_0$& $\Delta_0$
 &  $\chi^2$
\\
&Fig.~\ref{fig:SFLdens_gaps}&(meV)& $\overline{k_BT_c}$ &   $\overline{\chi^2(\Delta_0|\hat{k}_x\pm i\hat{k}_y|)}$    
\\
 \hline
$\Delta_0$          & b & 1.33(4)&1.95(5)   &  3.79 \\

$\Delta_0|\hat{k}_x \pm i\hat{k}_y|$ & c & 1.82(4)  &2.68(5) & 1 \\

$\Delta_0(1-\hat{k}_y^4)$ & d & 1.56(4)  &2.29(5) & 1.12   \\

$\Delta_0|\hat{k}_y|$ & e & 3.32(20) & 4.88(30) & 8.85 \\

$\Delta_0(1-\hat{k}_x^4-\hat{k}_y^4)$ & f & 2.61(10) & 3.84(14) & 3.29 \\

\hline
\hline
\end{tabular}
\end{center}
\end{table}

We are not aware of any work reporting the possible order parameter
symmetry in PrPt$_4$Ge$_{12}$. This issue was thoroughly studied
for another representative of the skutterudite family
PrOs$_4$Sb$_{12}$ \cite{Elbert03, Maki04}. Consequently, the
temperature dependence of the superfluid density  was analyzed by
assuming the following gap symmetries:
(1) the isotropic gap $\Delta(\theta, \phi) = \Delta_0$
(Fig.~\ref{fig:SFLdens_gaps}b),
(2) $\Delta(\theta, \phi) = \Delta_0|\hat{k}_x \pm i\hat{k}_y| =
\Delta_0 \sin \theta$ (Fig.~\ref{fig:SFLdens_gaps}c),
(3) $\Delta(\theta, \phi) = \Delta_0 (1 - \hat{k}_x^4 - \hat{k}_y^4)
= \Delta_0( 1-\sin^4\theta \cos^4\phi - \sin^4\theta \sin^4\phi)$
(Fig.~\ref{fig:SFLdens_gaps}d),
(4) $\Delta(\theta, \phi) = \Delta_0(1 - \hat{k}_y^4) = \Delta_0 (
1- \sin^4\theta \sin^4\phi)$ (Fig.~\ref{fig:SFLdens_gaps}e),
and (5) $\Delta(\theta, \phi) = \Delta_0 \hat{k}_y = \Delta_0 |\sin
\theta \sin \phi|$ (Fig.~\ref{fig:SFLdens_gaps}f).
Here $\Delta_0$ is the maximum value of the gap at
$T=0$, and  $\theta$ and $\phi$ are the polar and the azimuthal
coordinates in the $k$-space. The temperature dependence of the gap
$\Delta(\theta, \phi, t)=\Delta(\theta, \phi)\delta(t)$ was assumed
to follow the weak-coupling BCS prediction $\delta(t) =
\tanh\{1.82[1.018(t - 1)^{0.51}]\}$ \cite{Carrington03}. Note, $\delta(t)$
is practically independent of gap model \cite{Prozorov06review}.
Although the gaps described by (2) and (3) have
similar shape (see Figs.~\ref{fig:SFLdens_gaps}c and d), the $\Delta
= \Delta_0 (1 - \hat{k}_y^4)$ one corresponds to triplet pairing
\cite{Maki04} which is observed in PrOs$_4$Sb$_{12}$ \cite{Aoki03}.

The temperature dependence of the superfluid density was calculated
within the local (London) approximation 
by using the equations \cite{Prozorov06review}:
\begin{equation}
\rho_{^{aa}_{bb}} = 1-\frac{3}{4\pi T}\int
\sin^2\theta{\begin{pmatrix}{\cos^2\phi}\\
{\sin^2\phi}\end{pmatrix}}
\cosh^{-2}\left(\frac{\epsilon^2+\Delta^2}{2T}\right)d\epsilon d\phi
d\theta
\end{equation}
\begin{equation}
\rho_{cc} = 1-\frac{3}{2\pi T}\int \cos^2\theta \cos^2\phi
\cosh^{-2}\left(\frac{\epsilon^2+\Delta^2}{2T}\right)d\epsilon d\phi
d\theta
\end{equation}
Here, $\rho_{ii}$ ($i=a$, $b$, or $c$) is the component of the
superfluid density along $i$th principal axis. Since our experiments
were performed on a polycrystalline sample, we used the powder average
of the superfluid density:
\begin{equation}\label{eq:SFLdPowd}
\rho_s = (\sqrt{\rho_{aa}\rho_{bb}} +
\sqrt{\rho_{aa}\rho_{cc}} + \sqrt{\rho_{cc}\rho_{bb}})/3.
\end{equation}
The fit was made for the only free parameter $\Delta_0$.
The results of the fit are presented in Fig.~\ref{fig:SFLdens_gaps}a
and Table~\ref{tab:aSFLt}.
The goodness of fit was checked by using the $\chi^2$ criterium. The
values of $\chi^2$ normalized to that obtained for the best
$\Delta_0|\hat{k}_x \pm i\hat{k}_y|$ model are given in the last
column of Table~\ref{tab:aSFLt}.  It is obvious that the fits using
gaps of the form: $\Delta = \Delta_0 (1 - \hat{k}_x^4 - \hat{k}_y^4)$,
$\Delta = \Delta_0 \hat{k}_y$, and $\Delta = \Delta_0$ are very
poor. The smallest $\chi^2$ was obtained for the $\Delta = \Delta_0
|\hat{k}_x \pm i\hat{k}_y|$ model, while the $\Delta = \Delta_0 (1 -
\hat{k}_y^4)$ model results in $\simeq$12\% larger $\chi^2$ value.
The gap-to-$T_c$ ratios were estimated to be $\Delta_0/k_BT_c =
2.68(5)$ and  2.29(5) for the $\Delta_0|\hat{k}_x \pm i\hat{k}_y|$
and $\Delta_0(1 - \hat{k}_y^4)$ gap functions,
respectively. Both ratios are higher than the BCS weak
coupling value 1.76, thus suggesting that PrPt$_4$Ge$_{12}$ is a
moderately strong coupled superconductor. They are also
relatively close to $\Delta/k_BT_c = 2.35$ obtained from specific heat \cite{Gumeniuk08}.

To conclude, the PrPt$_4$Ge$_{12}$ skutterudite superconductor was
studied in $\mu$SR,  specific heat, and electric resistivity experiments. For
$T/T_c < 0.2$ a clear $T^3$ dependence of the electronic specific
heat term gives evidence that the gap function of PrPt$_4$Ge$_{12}$
has point nodes. Analysis of the $\mu$SR data was performed by exact
minimization of the Ginzburg-Landau free energy. The dependence of
the superfluid density  ($\rho_s$) on the magnetic field was found
to follow $\rho_s \propto 1-(B/B_{c2})^{0.25}$.
The temperature dependence of
$\rho_s(T)$ was found to be well described by two
models with axial symmetry  with point-like nodes in
the gap: $\Delta =\Delta_0| \hat{k}_x \pm i\hat{k}_y|$ and $\Delta =
\Delta_0(1 - \hat{k}_y^4)$ despite of cubic symmetry of crystal structure.
The maximum gap-to-$T_c$ ratios were
estimated to be $\Delta_0/k_BT_c = 2.68(5)$ and 2.29(5), respectively, in agreement
with 2.35 reported in Ref.~\onlinecite{Gumeniuk08} based on
specific heat results.

This work was performed at  the Swiss Muon Source (S$\mu$S), Paul
Scherrer Institut (PSI, Switzerland). We would like to acknowledge
E. H. Brandt for valuable discussions.

\end{document}